\documentclass[runningheads,a4paper]{llncs}

\usepackage{amssymb}
\usepackage{graphicx}
\usepackage{cite}
\usepackage{paralist}
\usepackage{todonotes}
\usepackage{times}

\usepackage{url}
\urldef{\mailsa}\path|[johan.linaker, hussan.munir, per.runeson, bjorn.regnell]@cs.lth.se|
\urldef{\mailsb}\path|claes.schrewelius@sonymobile.com|

\newcommand{\keywords}[1]{\par\addvspace\baselineskip
\noindent\keywordname\enspace\ignorespaces#1}

\begin{document}

\mainmatter  

\title{A Survey on the Perception of Innovation in a Large Product-focused Software Organization}

\titlerunning{A Survey on the Perception of Innovation in a Large Software Organization}

\author{Johan Lin{\aa}ker  \and Husan Munir  \and Per Runeson \and Bj{\"o}rn Regnell \and \\Claes Schrewelius}
\authorrunning{Johan Lin{\aa}ker et al.}

\institute{Software Engineering Research Group, Computer Science, Lund University\\
\mailsa}

%
%

\toctitle{Lecture Notes in Computer Science}
\tocauthor{Authors' Instructions}
\maketitle

\vspace{-15pt}

\begin{abstract}
\textbf{Context.} 
Innovation is promoted in companies to help them stay competitive. Four types of innovation are defined: product, process, business, and organizational. 
\textbf{Objective.} We want to understand the perception of the innovation concept in industry, and particularly how the innovation types relate to each other.
\textbf{Method.} We launched a survey at a branch of a multi-national corporation.
\textbf{Results.} From a qualitative analysis of the 229 responses, we see that the understanding of the innovation concept is somewhat narrow, and mostly related to product innovation. A majority of respondents indicate that product innovation triggers process, business, and organizational innovation, rather than vice versa. However, there is a complex interdependency between the types. We also identify challenges related to each of the types. 
\textbf{Conclusion.} Increasing awareness and knowledge of different types of innovation, may improve the innovation. Further, they cannot be handled one by one, but in their interdependent relations.
\end{abstract}

\vspace{-10pt}

\section{Introduction}
\label{sec:Introduction}
In recent years, the focus on innovation has increased in many lines of business. Novel products and services have always been important, while with an increasing pace of change, new technologies and market concepts being launched, with small vendors coming up and changing the scene in very short time, the need for continuous innovation is stressed in larger companies. Internet technologies for communication and distribution, and products and services primarily differentiated with respect to software, enables this shift by lowering the thresholds for new actors, and thereby threatening the position of existing ones.

Innovation is not only bringing new products to the market. The Organisation for Economic Co-operation and Development (OECD) Oslo manual~\cite{oecd2005}, which is used to guide national statistics collection on innovation, distinguishes between four categories of innovation, 
\begin{inparaenum}[i)]
\item product,
\item process,
\item marketing, and
\item organizational.
\end{inparaenum}
These categories are defined as follows: \emph{A product innovation is the introduction of a good or service that is new or significantly improved with respect to its characteristics or intended uses}~\cite[\S 156]{oecd2005}, while a \emph{process innovation is the implementation of a new or significantly improved production or delivery method}~\cite[\S 163]{oecd2005}. In the context of software engineering, we also count software development processes and practices as ``production" methods in the process innovation category. \emph{A marketing innovation is the implementation of a new marketing method involving significant changes in product design or packaging, product placement, product promotion or pricing}~\cite[\S 169]{oecd2005}. Note that this involves the whole concept of bringing a product or service to the market, a kind of innovation we have seen in the software and internet domain, for example, using information or advertising instead of money as a trade for services. Finally, an \emph{organisational innovation is the implementation of a new organisational method in the firm's business practices, workplace organisation or external relations}~\cite[\S 177]{oecd2005}. This is also prevalent in software, where for example open source software, outsourcing and offshoring significantly has changed the game in many lines of business.

Given these categories of innovation, we were interested in studying to what extent these were known and integrated in the culture of a large company, which is under rapid change, and where innovation is a key survival factor, due to the volatility of the market. In particular, we wanted to study the awareness of the innovation concepts, and the interplay between the four types of innovation; which types precedes the other? There is a similarity to the software process improvement trinity of people, process and technology, much discussed in the 1990's \cite{Humphrey89}. More specifically, this study formulates three research question:

\begin{enumerate}[RQ1]
\item What are the general perceptions of the term \emph{innovation}?
\item What relations are assumed between \emph{product} innovation and \emph{process, organizational} and \emph{marketing} innovation, respectively?
\item Which challenges exist with respect to the four types of innovation?
\end{enumerate}

To address the research questions we launched an internal online survey \cite{Fink2003} in a local branch of a multi-national corporation. The target population consisted of approximately 900 employees. On a global level the company employs approximately 5,000.

We found that the understanding of the innovation concept is somewhat narrow, and mostly related to product innovation. A majority of respondents indicate that product innovation triggers process, business, and organizational innovation, rather than vice versa. However, there is a complex interdependency between the types.

The paper is outlined as follows. In Section~\ref{sec:RelatedWork} we summarize empirical studies on people's attitudes to innovation in software engineering. Section~\ref{sec:Methodology} describes the methodology and design of the survey, as well as threats to validity and a characterization of the case company. In Section~\ref{sec:Results}, we report our findings from the survey, and analyze the data. Section~\ref{sec:Conclusion} concludes the paper.

\section{Related work}
\label{sec:RelatedWork}
Innovation related to information technology (IT) has become vital part of most organizations' success, primarily for two reasons: i) growing importance of innovation for organizational life, and ii) the introduction of IT into almost every business unit of organizations~\cite{fichmangoing2004}. Lee and Xia~\cite{leeability2005} addressed the process bottlenecks to innovation, where development teams are inefficient and reactive in most cases. Consequently, this causes problems with lack of support for business adaptions to shifting demands. Agile development seem to offer remedy to make the whole process more innovative for product development and help development teams to quickly deliver innovative, high quality solutions to an ever increasing demand of business innovation~\cite{highsmithagile2001}. 

On the other hand, research evidence~\cite{conboybeyond2011} also suggest that agile could also be a hindrance for product innovation. It creates barrier in transferring the ideas outside the team boundaries due to short iterations and feature backlog reduced the amount of time that teams could spent trying new things or sharing new ideas across different teams. Wnuk et al. ~\cite{Wnuk} also hinted the fact that existing requirements processes are designed to handle mature features and consequently, raises the question of process innovation by having a separate requirements engineering process to make room for innovative features (other than featured backlog) in the products.

Lund at al.~\cite{Lund2012} conducted a survey to explore the effects that reutilization have on innovation. Results revealed that standardization of process will free up time for innovation and most interestingly, routines are capable of having positive impact on occurrence of ideas and follow through on ideas. Furthermore, paring routines with openness to continuously improve the existing routines leverage positive effects on innovation. Therefore, take away from the study for managers is to take a look at existing routines with the spectacle of improving them, which will not only improve the efficiency but also the innovation aspect.

Moreover, another study was found where Harrison et al.~\cite{harisonapplying2010} conducted a survey with 170 Finnish software organizations to explore the impact of human capital on open innovation. Therefore, it can be used as an example where people are affecting the innovation activities in the organization. The study findings suggest that software companies with the larger academically educated staff are more likely to apply open innovation business strategies to accelerate their internal innovation process. The study further argued that this could be due the strong ties between communities and universities. 
Similarly, Nirjar~\cite{nirjaraccruing2013} also performed a survey with 121 software companies across India to explore the impact of workforce commitment on the innovation capability of the software enterprises. The study findings highlighted that the commitment of the managers of software firms can significantly enhance the innovation productivity by creating certain policies (i.e. open business model)~\cite{chesbrough_open_2003} and practices/processes.

\section{Methodology}
\label{sec:Methodology}
In this section we describe the surveyed company more thoroughly and elaborate on the survey design, analysis and threats to validity.

\subsection{About the company}
The company, which is a multi-national corporation with approximately 5,000 employees globally, develop embedded devices and the studied branch is focused on software development for communication hubs and additional connected devices in an internet of things (IoT) fashion. We consider the studied company a representative case~\cite{Runeson12Case} for similar ones, and hypothesize that the findings have a much broader generality than just this company. The studied branch of the company has 1,600 employees, of which 800 work on software development for the devices, and 100 work on connected devices. 

The company develops software in an agile fashion and uses software product line management (SPL)~\cite{Pohl2005}. The company has defined more than 20,000 features and system requirements across all the product lines. Considering the innovation aspect, the company is moving from a closed innovation model to an open innovation model~\cite{chesbrough_open_2003},  through the use of open source software to exploit the external resources to accelerate their innovation process. The open source solution, referred to as \emph{the platform}, is the base for their software product line projects and derived products. New projects on the product line typically entails 60 to 80 new features with an average of 12 new system requirements per feature. There are more than 20 to 25 development teams develop these features. 

\subsection{Survey design}
An internal online survey~\cite{Fink2003} was designed in collaboration between the researchers and company representatives, running an internal project, aimed at assessing and improving the innovation climate in the company. The questionnaire is composed of three major parts:

\begin{enumerate}
\item Factors that contribute to the innovation climate, based on Ekvall's scheme~\cite{Ekvall1996}. 
\item  Questions on the four types of innovation (product, process, organizational and marketing) and their relation, based on the OECD model \cite{oecd2005}.
\item Factors that hinder and help innovation, based on Jansen et al.'s  Open Software Enterprise model \cite{Jansen2012}.
\end{enumerate}

In addition to ranking and preference questions, the survey had fields for free input for most questions. The questions were defined in several iterations between researchers and company representatives, particularly to make the terminology of the survey understandable for the participants. Further, the survey was piloted to a small group of company representatives before the final launch.

One particular term was given certain care, namely \emph{marketing innovation}. The original definition is that  a \emph{marketing innovation is the implementation of a new marketing method involving significant changes in product design or packaging, product placement, product promotion or pricing}~\cite[\S 169]{oecd2005}. However, in the company context, the term was perceived to be only related to what the marketing department was responsible for, and thus too narrow. Therefore, we replaced the term with \emph{business innovation} and extended it to cover the process where the needs of the customers are captured as input for the product planning. This extends business innovation into the area of Requirements Engineering, which can be seen as a software engineering process, i.e. is covered by the process innovation definition. This area is therefore somewhat overlapped, but with the general distinction that high level capturing of requirements is mainly covered by the business innovation definition.

The survey was launched via the company intranet in October and November 2013 to about 900 employees via a census sampling, most of them being developers, of which 229 responded, i.e. a response rate of 25\%. 

\subsection{Survey analysis}
As the surveyed company is product-focused the surveys had a main focus on determining the level and perception of product innovation. Due to the attempt to address the more general innovation questions, the analysis focuses on three of the questions, connecting product innovation to process, business and organizational innovation. 

The respondents were asked to ``select the more likely scenario'' in the following questions:
\vspace{-1mm}
\begin{itemize}
\item The product innovation triggers the process innovation, or vice versa
\item The product innovation triggers the business innovation, or vice versa
\item The product innovation triggers the organizational innovation, or vice versa
\end{itemize}
\vspace{-1mm}

This gave an ordinal scale with two options to answer which makes any attempt of drawing conclusions limited, although a general pattern was observed, as shown in Figure~\ref{fig:InnovationTriggers}. The survey generated 469 free text comments. Except for the three earlier mentioned questions, comments were mainly gathered from four questions where the respondents were asked how innovative (s)he perceived the organization to be with respect to the four types of innovation.

Qualitative analysis with a thematic approach \cite{Cruzes14} was used to analyze the data, which was codified in up to three levels. Based on the codified data and the comments in general, perception of innovation concepts were analyzed (Subsection~\ref{sec:Perceptions}) and the connections between product innovation and process, business and organizational innovation, respectively were identified (Subsections~\ref{sec:ProdVsProc}--\ref{sec:ProdVsOrga}). Further on, based on the themes and comments in general, challenges were then identified and generalized in regards to the four types of innovations (Subsections~\ref{sec:ProdChallenges}--\ref{sec:OrgaChallenges}). 

\subsection{Threats to validity}
The construct validity~\cite{kitchenham08}, refers to whether the survey measured what it was intended to. This can be addressed through e.g. pilot studies, which was performed before the official launch. Further on, the questions were developed iteratively and based on established literature.

In regards to the analysis, a threat to the construct validity is the risk of researcher subjectivity as the first author performed the mapping and main analysis. This was addressed by having the second and third authors perform their own individual analysis of the data, and could compare their findings with that of the first author.

External validity regards whether the results be generalized to outside of the surveyed sample~\cite{Runeson12Case}. In this paper, we analyze the questions, which can be published from the company's confidentiality perspective. Thus, we do not focus on their perceived current innovation status, but rather on the general understanding of innovation factors and their relations. Thereby, we also focus on the most generalizable aspects, which we hypothesize are valid for other companies of similar characteristic to the studied one, as a representative case~\cite{Runeson12Case}.

A surveys reliability~\cite{kitchenham08} concerns whether the same results can be obtained if the survey process was repeated. As the sample was obtained through a census sampling frame and had a response rate of 25\% we regard this optimistically. Although, this cannot be strengthened until follow-up surveys are performed. This is something that will be done in the future as the company wants to measure how the internal perception of innovation develops over time.

\section{Results}
\label{sec:Results}
\begin{figure}[t]
\begin{center}
\includegraphics[scale=0.6]{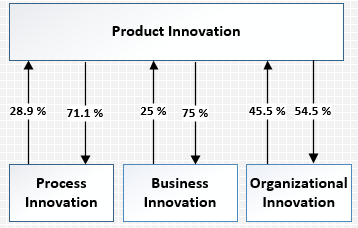}
\caption{Triggering relation between the four types of innovation: product, process, business and
organizational. Percentage value shows the share of respondents that select X $\rightarrow$ Y as the
most likely scenario.}
\label{fig:InnovationTriggers}
\end{center}
\vspace{-25pt}
\end{figure}

In this section we present our findings from the qualitative analysis of the survey responses. First the general perceptions of innovation is presented based on survey responses in \ref{fig:InnovationTriggers}. Then connections between product innovation and process, business and organizational innovation is presented respectively. Direction of arrows show the innovation type triggering the leading innovation (see fig. \ref{fig:InnovationTriggers}). For instance, the arrow from process innovation to product innovation shows that 28.9\% respondents think that process innovation leads to product innovation. Similarly, the arrow from product innovation to process innovation suggest that 71.1\% respondents think that product innovation lead to process innovation and the same arrow pattern applies for other innovation types. Finally, the challenges identified in regards to each innovation type is listed. As the types of innovation relate to each other, the challenges are structured accruing to the type where it relates the most, although a challenge may affect more

\subsection{Perceptions of innovation}
\label{sec:Perceptions}
Although not general, it was observed among the comments that some had trouble relating to the term innovation as such. The borderline between when something goes from being an improvement or common functionality to an innovation is fluid. 
\emph{``I recognize that [company] does this often [\ldots] But I'm not sure if it's really innovative or just mindless changes.''}

Some respondents consider innovation as part of their everyday work, while others are a bit more unclear on the distinction between their everyday work and innovative activities, or just creativity as a process.
\emph{``As a designer the largest part of the task when bringing forward is to be creative. However there is a difference between being creative and being innovative.''}

A reason could be unawareness of what the company counts as innovations and examples of different types of innovations. \emph{``I don't know much about the innovations that we do. I didn't know about the [example feature] for instance''}. 

Some may not be aware of what they do could actually count as an innovative activity. \emph{``I work with support systems and not product development. Some part of the time goes into improving how we produce products.''}

Further on, some believed that they were not able to perform any innovative activities as it was not a part of their work description or role. A tester expressed how he was not able to innovate as he assumed this was a task dedicated to developers. Another tester reasoned similarly. \emph{``Working with testing so not much improvement in the product besides some ideas that pops up occasionally.''} 

This thinking was present on a general level in connection to all of the four types of innovation. As mentioned, this could be due to that the awareness is limited of how and where they can innovate. A better understanding needs to be achieved for the different types of innovations and how these interplay. \emph{``Most of all, I would say that I have only minor insight and understanding of this field [of organizational innovation].''}

A consequence may be that some believe innovation is not possible. \emph{``I don't think it is possible to be innovative in this area [organizational innovation].''}

Apart from spreading awareness and knowledge, another important factor that needs consideration is the mindset.
\emph{``Since I'm not involved in this part of our business then it's not in my mindset, but when you now mentioned it I will take it into my consideration of innovation.''}. 

\subsection{Product innovation vs Process innovation}
\label{sec:ProdVsProc}
On the question whether product innovation triggers process innovation, or the other way around, 71 percent answered the former (see fig.\ref{fig:InnovationTriggers}). Although the percentage points in one direction, it is clear from the free text answers that this question is more complex than so. 

Processes can be strict and complex, creating overhead and distraction, occupying time that could have been focused on creative thinking, as pointed out by a respondent.
\emph{``If the development process is driven as a rigid framework that is complex and difficult to understand who decides what and why, then you do not get in the dynamics of ideas.''}

This is also identified as a challenge of process complexity in Subsection~\ref{sec:ProcChallenges}. Although processes can force a static frame on employees, it can help to bring structure to the innovation process and thereby still encourage innovation and creative thinking. \emph{``\ldots well defined and established processes leads to innovative products.''}

Another challenge is idea tracing and execution uncertainty (see Subsection~\ref{sec:ProdChallenges}), which is an area where we hypothesize that well-designed processes can help to clarify what happens to ideas and the roadmap for how innovations can be pushed through. Similarly, processes can also help to increase the awareness of the product scope and the innovation strategies in the organization. 

Process innovation may help the organization become more efficient and reduce waste as can be interpreted by the OECD definition~\cite{oecd2005} and as pointed out by a respondent: 
\emph{``\ldots process innovation improve performance, simplifies and speeds-up development process - thus allowing to have more resources in true product innovation''}. This aligns with the area of Software Process Improvement~\cite{harter2000}, which includes possible implications from new or improved tools and techniques. As put by another respondent:
\emph{``\ldots We need to have the proper techniques, equipment and SW in order to develop new and improved products.''}

The resources made available can be defined as freed-up budget-hours, which can be used for other purposes, such as time dedicated to activities focused on rendering product innovation. An organizational and cultural challenge in this case is to actually make this dedication which demands a committed management.
\emph{``The process innovations are often meant to make development faster with more quality, but I'm not sure the gained resources are spent on product innovation.''} 

Beneficial factors from a process change, other than freed up resources, may also include an increase in performance and quality as confirmed by the respondents. Although, it is a matter of definition how software quality relate to product innovation~\cite{prahalad1998}, this will hopefully render in a better product offering which further down the release ladder may prove to be a trigger of future product innovations.

Hence, by innovating and improving the processes in the correct way and dedicating the freed up resources to product innovation, process innovation can be seen as a trigger for product innovation. This is in line with findings by Lund and Magnusson~\cite{Lund2012}. On the other hand, processes are not decoupled from the products. There needs to be an awareness of product roadmaps and an adaptive mindset as some processes may require continuous tailoring as a consequence. 
\emph{``I think the general mindset is "keeping the eye on the prize", you see the upcoming releases in the horizon and you adjust the process to meet those releases.''}

The need to adapt is not a simple task and requires both resources and dedication. Keeping pace with new features and products can be very demanding for an organization as pointed out by the respondents. Process changes needs to be quickly adopted for the organization not to fall behind or get confused, as described in the process innovation challenges (Subsection~\ref{sec:ProcChallenges}).

Just as new products may create a demand for new processes and tools, they can also be an inspiration for new techniques and solutions. \emph{``On the other hand, new products can also inspire new techniques and HW/SW solutions.''}.

\subsection{Product innovation vs Business innovation}
\label{sec:ProdVsBusi}

On the question whether product innovation triggers business innovation, or the other way around, 75 percent answered the former (see fig.\ref{fig:InnovationTriggers}). As with the previous question, although there is a clear majority in one direction, this does not give the complete answer.

Some see product innovation as the driver with respect to business innovation due to that \emph{``Innovative products are a great source for new business opportunities and marketing''}. Innovative features affects which consumer groups that should be targeted, and in effect which marketing channels that can be used. 
The nature of the innovative features also has implications on how the marketing message can be phrased and communicated. 
From this point of view, the products both enable and set a demand for a continuous business innovation that can adapt to changing functionality and feature sets. A good product as foundation, can even be seen as a source of inspiration to excel business innovation as hinted by the following respondent.
\emph{``I think everything starts with the product. If you are a company with "Wow!"-products then the rest will come. A consumer will see through (eventually) if the company is only selling a mediocre product but have brilliant marketing. However, if we have good products, it will be more motivating bringing it to the market, which will inspire us to excel also in business innovation''}

From the other perspective, innovative marketing may be a requirement for what otherwise would be considered a normal product. Competitive products, which are technically inferior, may very well prove more popular compared to a technically superior product, due to the awareness and visibility towards the customers, as identified by the respondents. Business innovation can create the hype needed to tell about what the innovative features are, how they differentiate and how they fit in the customers' context. However, as pointed out by the previous quote, if the product does not fill the expectations, innovative marketing will not be a viable solution in the long run.

New innovative ways are continuously needed to keep pace and capture the demands from the existing and emerging customer channels, e.g. through end-user feedback~\cite{Bano2014}. An awareness of what needs the customers have today and will have tomorrow, is an important input from business and marketing to push the product innovations forward in the right directions. 
\emph{``Because business innovation brings in new experience directly from market, new demands and requirements and thus giving a product a right direction''}

This creates a challenge for the organization in terms of synchronization. The view of what features are to be considered game-changers and prioritized in the release planning process~\cite{carlshamre2002}, may prove troublesome due to internal communication gaps between marketing and product development~\cite{Karlsson2007}, which may lead to wrong features being promoted as a consequence. 
\emph{``Scope/product planning, business side and development [should be] in sync regarding both our innovation initiative [\ldots] and how to drive innovations all the way to product.''}

As explained, there is a dual sided relationship. There is a dependency going in both directions where one can trigger the other. One respondent provided a concrete example which summarizes the relationship.
\emph{``It is pretty much both. Look at the music and film business which has invented new ways of marketing and distribution, but I believe the wish of distribute TV via satellite has created new products for making it possible and to get paid for it. Then again we have the Google glasses. Right now they are cool, but not very useful until we find a useful feature for them and that itself will create a business for them.''}

\subsection{Product innovation vs Organizational innovation}
\label{sec:ProdVsOrga}

On the question whether product innovation triggers organizational innovation, or the other way around, 55 percent answered the former (see fig.\ref{fig:InnovationTriggers}). Opposed to the previous questions, this was not as clear majority for the product innovation centric view. 

Improving and innovating the way in which a company collaborates and interacts with external parties and stakeholder, can trigger product innovations in several ways. Application of open innovation business strategies is one way to accelerate their internal innovation process~\cite{harisonapplying2010}. Crowdsourcing ideas, engaging in Open Source communities, welcoming third-party developers, acquiring promising startups and starting joint-ventures or ecosystems are a couple of activities that falls into the open innovation paradigm originally defined by Chesbrough~\cite{chesbrough_open_2003}, that may render in new product innovations.

Creating a more innovative organizational environment with committed employees is another way that can lead to more product innovations~\cite{nirjaraccruing2013}, as described by a respondent:
\emph{``With a flexible and happy organization that makes people get looser boundaries I believe we can get a more innovative climate''}
Bringing people from different backgrounds and functional areas creates diversity and enables for new discussion to arise and to discuss ideas from new angles~\cite{calantone2002, koc2007}, or as put by the following respondent: 
\emph{``Connecting colleagues which hadn't possibility to communicate before allows to discuss more problems and ideas.''}. Calantone et al.~\cite{calantone2002} adds that this cross-functional integration also allows for the employees to evolve their skills by learning and sharing knowledge amongst each other, which is important for product development.

This connects to a need for a general awareness of what has been done, and what is being worked on. 
\emph{``\ldots more often than not these innovations are "hidden" in small segments of the company, not actively promoted and spread (and that's both good and bad, many projects dies when they need to become too big).''}
By communicating items such as features, functionality, experienced problems and related solution across internal borders, cross-functional views can be established more automatically. A solution in one project may turn out to solve the same issue or create new ideas in another project, which could either be considered a process or a product innovation. This relates to the concept of inner source~\cite{Linaaker2014} and how it can help organizations work more open and cross-functional, and in the end become more innovative~\cite{morgan2011}.

Organizational barriers and communication issues is another area, where organizational innovation may trigger product innovation in the long term perspective. When products or processes stretch over multiple business units or projects, this can create room for bureaucracy, different prioritization schemes, culture and politics, to mention a few factors~\cite{koc2007}. 
\emph{``Some sections within the company are quite innovative, but when it comes to cross-functional agreements and alignment, there always seems to be a resistance to change and adapt to new ways of working and safeguarding what seems to the best for "me/my team" is more important than what's best for the company.''}

Pushing through and spreading an idea across these borders require a high level of internal permeability.
\emph{``Organization organized for better collaboration (=no filtering, no proxies, smaller proximity, time zone, etc\ldots) is more likely to produce more innovative ideas. Layering, direct reporting, micro management, and similar old-school practices are killing innovation.''}

Looking from the other perspective, new product innovations will create new demands and implications which will give rise for possibilities and triggers for organizational innovation~\cite{calantone2002}. 
\emph{``New and exciting products means we have to adapt how we work to support these in the best-possible, not only from an engineering or software perspective, but for example from the launch projects etc.''}

As has been discussed in regards to previous sections on the matter of product innovation versus process and business innovation, there exists a dual relationship here as well as exemplified by the response:
\emph{``Organizational innovation increases our capability to handle new and complex tasks. Innovative products will require us to handle new or more complex tasks and without room for growth, product innovation will fizzle.''}

\subsection{Product innovation challenges}
\label{sec:ProdChallenges}
In the responses, several aspects were mentioned as challenges to the product innovation.

\begin{inparaenum}[a)]
\item \emph{Idea tracing and execution uncertainty} -- Even though there may be a rich pool of innovative ideas being produced and a general will to contribute, it is important to maintain and support it. Knowledge and awareness of what happens to ideas contributed to the innovation development process is important for the contributors to feel that they are taken seriously and that it is worth to continue contributing, which in turn gives an increased innovation capacity for the company~\cite{koc2007}. When the ideas come bottom-up there needs to be a feedback loop top-down that stimulates this need of information as confirmed by Koc and Ceylan~\cite{koc2007b}, and Wnuk et al.~\cite{Wnuk}. 

\item \emph{Short term perspective} -- By having a narrowed foresight, release planning tend to prioritize non-unique features which renders in low diversity in the product range, thus making the company being a follower of competitors rather than a leader. A longer time perspective needs to be integrated into the company culture, together with a positive mindset for game changers and innovative features to be created. 

\item \emph{Product scope and innovation strategy} -- Uncertainty about the product roadmap and feature scope leads to risks that the creative minds of the company are misdirected. A common and established innovation strategy can help defining the product scope and frame where ideas are needed suggested by Koc and Ceylan~\cite{koc2007b}, and  Wnuk et al.~\cite{Wnuk}.

\item \emph{Limiting environment and mindset} -- Soft factors such as employees feeling that they can have a free mindset and share ideas openly is important for an innovative environment. It must be okay to test new ideas, but also to fail. These are factors, triggered by Ekvall's innovation climate model~\cite{Ekvall1996}.

\item \emph{Restriction by external stakeholders} -- A commercial product company can have many stakeholders, some not being the end customer. This may include distributors and service providers further down the value chain, adding value and modifications to the product before they reach the final buyers. These stakeholders put requirements that may prevent and limit the feature scope possible to address. This filter risks to kill ideas inside the company and ignore needs, both identified and unidentified, from the end customers. This challenge is in line with Conboy and Morgan's findings~\cite{conboybeyond2011}.

\item \emph{Limited time for innovation activities} -- Tight project budgets and short deadlines are two factors that can restrict time available for idea creation. Developers usually have pet projects and ideas they would like to work on, some even dedicate their spare time for this purpose. By allowing the time, this can prove a valuable source of product innovation as suggested by Conboy and Morgan~\cite{conboybeyond2011}.


\item \emph{Cross-functional resources} -- Bringing new people together creates new product ideas and can boost innovation development. Cross-functional labs-sections and dedicated innovation team are two examples suggested by Conboy and Morgan~\cite{conboybeyond2011}, and Koc~\cite{koc2007}.
\end{inparaenum}

\subsection{Process innovation challenges}
\label{sec:ProcChallenges}
This section presents the challenges, directly related to process innovations.

\begin{inparaenum}[a)]
\item \emph{Process change too slow} -- The introduction of a new process may be cumbersome for several reasons, with the effect that the changes are implemented slowly. This can cause confusion for employees being caught between two states -- before and after the change -- and also result in an unsynchronized organization as different parts may adapt faster than others. 

\item \emph{Process change too often} -- Another issue with respect to process change is that they may happen too often. This can be a cause effect relationship with an adoption process, as old processes risk being outdated once introduced if done in a too slow and inefficient manner. When the environment changes, for example technology and dependencies towards partner's progress, so does the requirements on the internal tools and processes have to change at the same pace. This can also relate to organizational innovation. 

\item \emph{Process change top down} -- Problems can arise when a process is introduced top-down instead of bottom-up. Managers may not always know what is the most efficient way to work compared to those actually performing the work. This challenge is also in line with the findings of Qin~\cite{qin2007}, and Wnuk et al.~\cite{Wnuk}.
\end{inparaenum}

\subsection{Business innovation challenges}
\label{sec:BusiChallenges}
Challenges related to business innovation are about alignment with the market and end users.

\begin{inparaenum}[a)]
\item \emph{Reaching the end-customers} -- When there are layers between the producer and end-customer, for example, distributors and service providers, promotion of new ideas and product innovations to end-customers gets complicated. As technology and social habits evolve, new innovative ways are needed to keep pace with the different forums for communication used by the end-customers of today and tomorrow. Examples of such phenomena are software ecosystems \cite{WnukEcosystem14}.

\item \emph{Product and marketing synchronization} -- The views on what the top innovative features are may differ between different parts of the company. A misalignment like this can create confusion between marketing and product development. This could render in the wrong features being promoted. The suggested needs of the end customers should be communicated and synchronized to all relevant parts of the organization, e.g. product planning, marketing and development. 
\end{inparaenum}

\subsection{Organizational innovation challenges}
\label{sec:OrgaChallenges}
Organizational innovation challenges relate to collaboration, communication and change. 

\begin{inparaenum}[a)]
\item \emph{Closed organizational borders} -- If the organization is too introvert and closed, opportunities, possible collaborations, sources of ideas and other possible inputs to their internal innovation process might be missed. By opening up the company borders for external collaboration and influence, new possibilities can arise both in regards to new innovations and markets, as described by the Open Innovation paradigm~\cite{chesbrough_open_2003}.

\item \emph{Intra organizational collaboration} -- Barriers and layers can prevent otherwise prosperous and potential collaborations between business units in organizations. Examples may be different sub-priorities of features between projects and multiple number of mangers creating a complex and bureaucratic hierarchy~as identified among the respondents and confirmed by Koc~\cite{koc2007}. These are related to what Bjarnason et al refer to as ``gaps'' \cite{BjarnasonWR11}. Koc further points out that such cross-functional integration demands a high level of coordination, otherwise it will rather have a negative impact on the product innovation.

\item \emph{Intra organizational learning} -- Unawareness of what has been done in other parts of the company can create inefficiency and missed possibilities. In regards to process innovation, tools, technologies and processes from one part may prove its self superior or complementary to those used in other parts. And in regards to product innovation, a commoditized good or service from one business unit may turn out as innovative if added to the value proposition in another business unit's product chain. This is a challenge in-common with inner source~\cite{Linaaker2014}, but also one of the ways in how it can help organizations become more innovative by using it as a type of intra-organizational open innovation~\cite{morgan2011}.
\end{inparaenum}

\section{Conclusions}
\label{sec:Conclusion}
The view on what innovation is and where it can be performed is a diversified topic. OECD~\cite{oecd2005} differentiates between four types: product, process, market and organizational innovation. These were adopted in the survey on which this paper is based on, with a redefinition of market innovation into business innovation. The original definitions are general and applicable on a multiple number of fields. This paper puts them in the context of software engineering characterized by the opinions of people involved in different levels of a large software development organization.

The perception of the term \emph{innovation}, to answer the first research question (See \textbf{RQ1}, Section~\ref{sec:Introduction}), is diversified. Even though it is not general, some had trouble relating to the term innovation as such and when a feature or certain work can be classified accordingly. Some believed that they were not able to perform any innovative activities as it was not a part of their work description or role, which was present in connection to all of the four types of innovation. Apart from awareness and knowledge, another important factor that also needs consideration is the mindset of the employees that innovation is possible and something that they can help to create.

The different types cannot be considered isolated or decoupled which answers the second research question (See \textbf{RQ2}, Section~\ref{sec:Introduction}). Connections between product innovation and process, business and organizational innovation exists in both directions. Introduction of product innovations creates demand and possibilities for processes, marketing and organization to adapt and optimize as the conditions has been changed. Interdependencies may require tailoring being done, either as a direct consequence or as a side effect. On the other way around, introduction of a process, business or organizational innovation can change the environment and conditions for how product development is being done. Inputs such as new technologies, ideas, resources and know-how are example factors which can be considered a cause behind a product innovation effect. Open innovation could be classified as an organizational innovation that can render inputs to the internal innovation process~\cite{chesbrough_open_2003}. 

Challenges correlated to the different innovation types were also identified, with respect to the third research question (See \textbf{RQ3}, Section~\ref{sec:Introduction}). These give a context to the term of innovation that covers parts other than the more normal conception of innovation in regards to just products. Some challenges may target more than one type of innovation, e.g. internal communication which can cause issues for introduction on new processes and organizations as well as hinder ideas to be spread and discussed. 

For future research it would be interesting with studies confirming and exemplifying the connections described, for example how process innovation could trigger product innovation. An anticipated challenge will be to trace a cause effect relationship and connecting the two areas. Another area also includes confirming the challenges identified, and further characterizing the innovation types from a software engineering perspective.

%
%
\bibliographystyle{plain}
\bibliography{bibliography}
\end{document}